\documentclass[letterpaper]{article}
\usepackage[T1]{fontenc}

\usepackage{geometry}
\usepackage{setspace}
\usepackage{amsmath,amssymb}
\usepackage[style = nature, articletitle=true, doi=false]{biblatex}
\let\cite\supercite

\usepackage{graphicx}

\usepackage{float}
\usepackage{placeins}
\usepackage{booktabs}
\usepackage{multirow}
\usepackage{caption}
\newcommand{\KMD}{K_{\mathrm{I}}^{\mathrm{MD}}}
\newcommand{\Kapp}{K_{\mathrm{I}}^{\mathrm{app}}}
\newcommand{\KIG}{K_{\mathrm{IG}}}
\newcommand{\KIe}{K_{\mathrm{Ie}}}
\newcommand{\muH}{\mu_{\mathrm{H}}}
\newcommand{\cH}{c_{\mathrm{H}}}

\title{Machine-learned atomistic simulations reveal the basis of hydrogen-induced crack-plane transition in $\alpha$-Fe}

\usepackage{authblk} 

\author[1]{Jiaqin Xu*}
\author[1]{Zhiqiang Zhao*}
\author[2]{Kazuma Ito}
\author[1]{Shuhei Shinzato}
\author[1]{Fanshun Meng}
\author[1]{Shihao Zhang}
\author[1]{Shigenobu Ogata*}

\affil[1]{Department of Mechanical Science and Bioengineering, Graduate School of Engineering Science, The University of Osaka, Osaka 5608531, Japan}
\affil[2]{Advanced Technology Research Laboratories, Nippon Steel Corporation, Chiba 2938511, Japan}

\date{*Email: xujiaqin@tsme.me.es.osaka-u.ac.jp; zhiqiang.zhao@tsme.me.es.osaka-u.ac.jp; ogata.shigenobu.es@osaka-u.ac.jp}

\begin{document}

\onehalfspacing
\maketitle

\doublespacing
\begin{abstract}
Hydrogen-related fracture in body-centered cubic Fe and ferritic steels often appears as transgranular quasi-cleavage rather than purely intergranular failure, especially at low to moderate hydrogen contents.
Fractography has suggested that hydrogen may change the dominant cleavage faceting from $\{100\}$ toward $\{110\}$, but atomic-scale evidence for this possible crack-plane transition remains unclear.
Here we construct an efficient neural-network potential for $\alpha$-Fe/H and combine large-scale, three-dimensional molecular dynamics with grand-canonical Monte Carlo (GCMC), allowing the near-tip crack-surface region and crack tip within a defined GCMC domain to exchange hydrogen with a reservoir at fixed chemical potential.
A comparison of four crack systems identifies the controlling response: $(100)[010]$, $(100)[011]$, and $(110)[001]$ remain cleavage-dominated, whereas the $(110)[1\bar{1}0]$ crack changes from dislocation emission in pure Fe to cleavage under hydrogen charging.
The energetic origin is twofold.
Hydrogen lowers the Griffith cleavage threshold of the $\{110\}$ cleavage-plane family more strongly than that of $\{100\}$, and, for the controlling crack, a Rice-type energetic descriptor indicates that the surface-energy-controlled cleavage resistance decreases faster than the unstable-stacking-fault-controlled emission resistance, consistent with a weakened dislocation-emission shield.
These results provide a thermodynamically consistent atomistic basis for a hydrogen-induced transgranular crack-plane transition in Fe.
\end{abstract}

\begin{quotation}
\end{quotation}

\clearpage

\section{Introduction}
\label{sec:introduction}
Hydrogen embrittlement remains a central barrier to the reliable use of Fe and steels in structural components exposed to hydrogen-bearing environments.
Although hydrogen-assisted fracture is often discussed in terms of a reduction in toughness, the fracture surface itself can also change.
In body-centered cubic (bcc) steels, hydrogen-related fracture at comparatively low or moderate hydrogen contents frequently appears as transgranular quasi-cleavage, whereas intergranular fracture becomes prominent in other regimes of chemistry, strength, stress state, and hydrogen content \cite{bernstein1970hydrogen,ogawa2019hydrogen,okada2022origin,okada2023characteristics}.
Recent crystallographic analyses of hydrogen-related quasi-cleavage in bcc ferritic and martensitic steels have further suggested a change in dominant facets from $\{100\}$ toward $\{110\}$ facets \cite{okada2022origin,okada2022effect,bestautte2023investigation}.
This possible crack-plane transition is important both scientifically and technologically. 
Scientifically, it implies that hydrogen does not merely reduce fracture resistance; it reorders both the competition among cleavage paths and the competition between cleavage and dislocation emission at the crack tip. 
Technologically, a change in the active transgranular cleavage plane alters crack-path selection, local shielding, and the transferability of toughness data across hydrogen exposure conditions, all of which are central to fracture assessment and alloy design for hydrogen-bearing infrastructure.

The key unresolved question is therefore not only whether hydrogen is present near a crack, but which hydrogen state is thermodynamically selected at crack surfaces and crack tips under load, and how that state changes the local fracture criteria.
Experiments can identify crack paths and fracture-surface facets \cite{bernstein1970hydrogen,okada2022origin,okada2022effect,bestautte2023investigation}, but the hydrogen occupation that controls crack-tip energetics is not directly observable with atomic resolution.
A predictive atomistic description must therefore treat the near-tip crack surfaces and crack-tip region as chemically open defects coupled to a hydrogen reservoir.
In that description, the near-tip hydrogen population is an outcome determined by the chemical potential and loading state, not an assumed inventory or a kinetic hydrogen-delivery rate.

Atomistic simulation can in principle provide this information, but two limitations have prevented a decisive test of the proposed $\{100\}$ to $\{110\}$ transition.
First, the problem is intrinsically three-dimensional: the crack front must be long enough to allow curved dislocation emission, cleavage embryos, and mixed events to compete along the front.
Second, because interstitial hydrogen is highly mobile in bcc Fe at room temperature, a fixed-chemical-potential treatment is more appropriate than prescribing a fixed number of pre-charged atoms when the near-tip hydrogen population is expected to adjust to the evolving crack-tip state.
Earlier density functional theory (DFT) studies provide reliable energetics but are too costly for this scale, whereas conventional empirical potentials can misrepresent even the fracture behavior of pure bcc Fe \cite{moller2014comparative,suzudo2022cleavages}.
Recent machine-learning interatomic potentials have greatly improved the fidelity of atomistic fracture simulations \cite{suzudo2022cleavages,zhang2023atomistic,meng2021general,zhang2024highly}; however, much of the existing literature still relies on thin two-dimensional crack models and/or fixed hydrogen contents, which suppresses three-dimensional crack-front competition and prevents comparison under a common thermodynamic hydrogen condition \cite{song2011nanoscale,song2013atomic}.

Here we address these gaps by developing a high-efficiency neuroevolution potential (NEP) for the $\alpha$-Fe/H system and using it in large-scale, three-dimensional, constant-chemical-potential crack simulations.
The approach allows four competing crack systems, $(100)[010]$, $(100)[011]$, $(110)[001]$, and $(110)[1\bar{1}0]$, to be compared under the same hydrogen reservoir condition.
This comparison first identifies the controlling crack system directly: among the four systems, only $(110)[1\bar{1}0]$ changes from dislocation emission in pure Fe to cleavage under hydrogen charging.
We then show that the proposed crack-plane transition is supported by two energetic signatures: hydrogen inverts the cleavage preference between crystallographic cleavage-plane families by reducing the $\{110\}$ Griffith threshold more strongly than the $\{100\}$ threshold, and a Rice-type energetic descriptor shows that hydrogen shifts the cleavage-emission balance toward cleavage for the controlling $(110)[1\bar{1}0]$ crack.
Together, these results provide a physical basis for a hydrogen-induced transgranular crack-plane transition in $\alpha$-Fe.

\section{$\alpha$-Fe/H neural-network potential}
\label{sec:nep_main}
Resolving hydrogen-dependent fracture in three-dimensional crack models requires an interatomic model that is both chemically accurate and fast enough for million-atom simulations.
We therefore constructed a neural-network interatomic potential for the $\alpha$-Fe/H binary system within the NEP framework implemented in GPUMD \cite{fan2022gpumd,xu2025gpumd}.
NEP combines high-dimensional local structural descriptors with a compact neural-network regression model, providing near-DFT accuracy at a computational cost suitable for large-scale molecular dynamics.
The mathematical form of the descriptors and the training protocol are described in Methods and Supporting Information Section~S1; here we focus on the validation results that are most relevant to fracture.

\begin{figure}[htbp]
  \centering
  \includegraphics[width=1.0\textwidth]{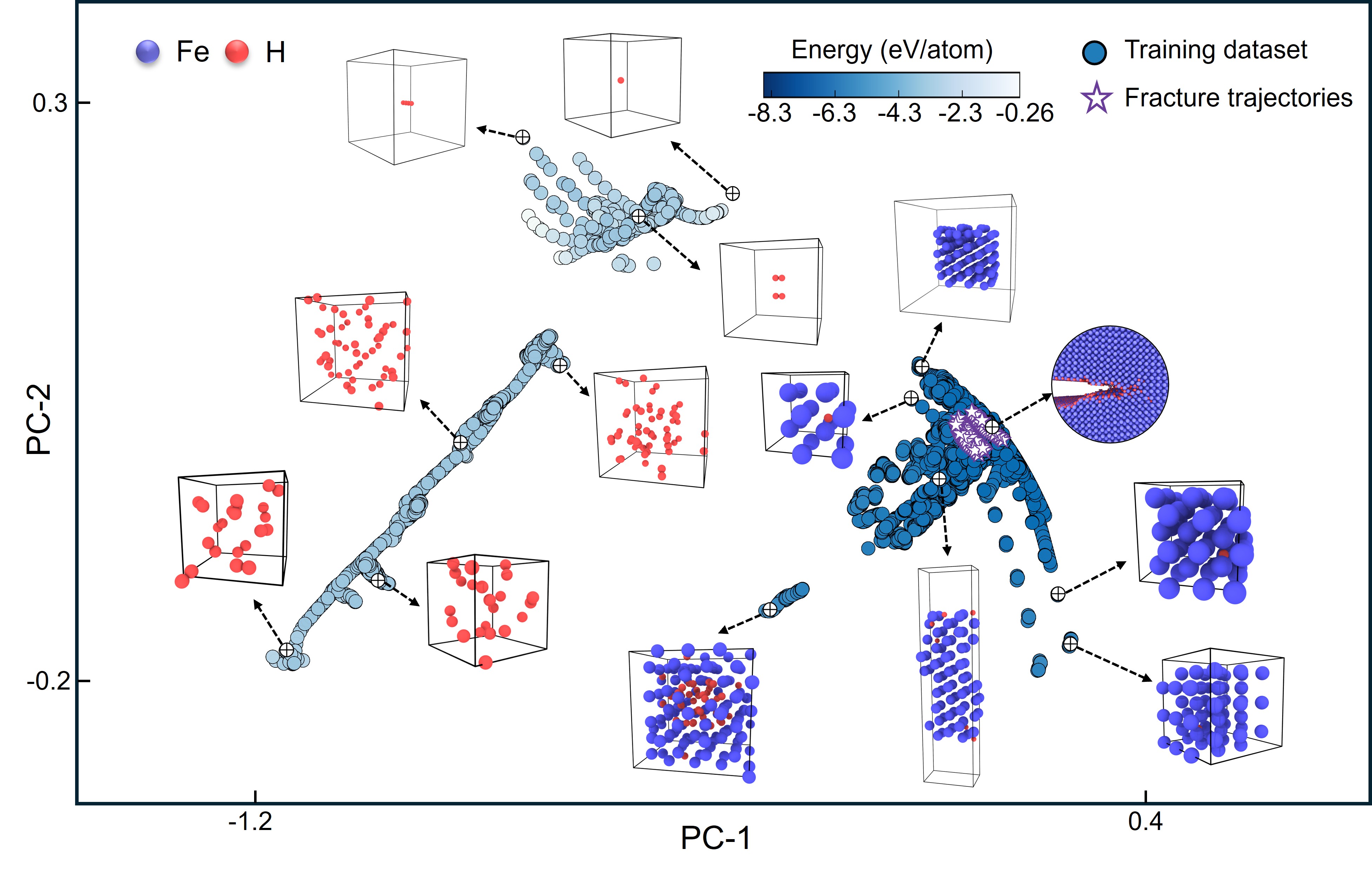}
  \caption{Training-data coverage for the $\alpha$-Fe/H NEP. 
           The 21,926 training structures are projected onto a two-dimensional principal-component (PC) subspace of the descriptor vector. 
           Color denotes atomization energy. 
           Representative structures show that the dataset spans hydrogen-only, Fe-rich, defect, surface, and fracture-related environments. 
           Purple stars indicate local environments sampled along the fracture trajectories; their location within the training distribution supports the transferability of the potential to the crack simulations.
           }
  \label{fig:nep_pca}
\end{figure}

The training dataset contains 21,926 DFT-labeled structures drawn from a broad range of Fe and Fe-H environments, including bulk phases, surfaces, vacancies, grain boundaries, dislocations, hydrogen in bulk and defects, and H-H interactions in Fe and vacuum \cite{meng2021general}.
Because the main fracture simulations sample hydrogen-decorated crack surfaces, crack tips, and emitted-dislocation embryos, the relevant validation question is whether those local environments remain within the domain represented by the training data.
The principal-component projection in Fig.~\ref{fig:nep_pca} shows that the fracture trajectories sampled in the present simulations remain embedded within the training-data distribution.
This coverage is important because crack propagation samples highly distorted local environments near surfaces, crack tips, emitted dislocations, and hydrogen-decorated defects.

\begin{figure}[htbp]
  \centering
  \includegraphics[width=1.0\textwidth]{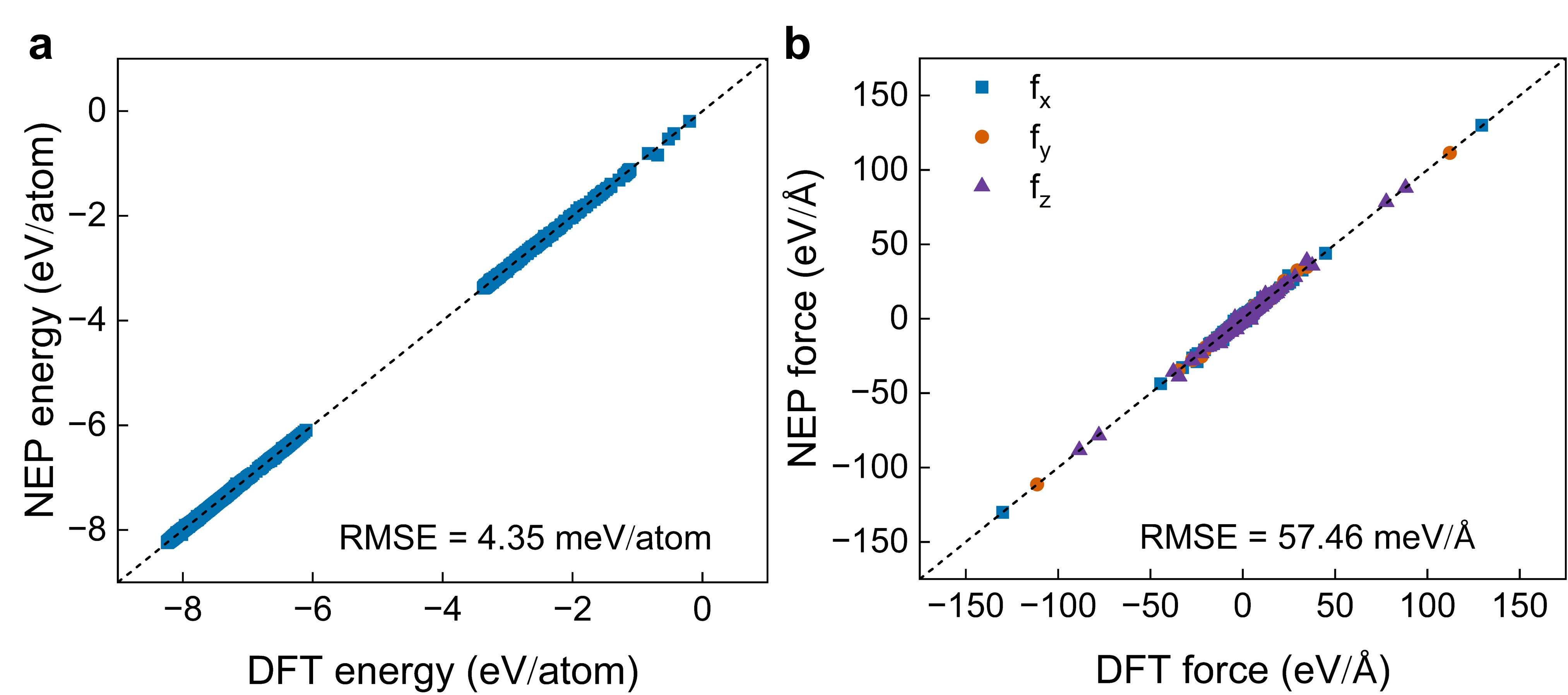}
  \caption{Accuracy of the $\alpha$-Fe/H NEP. 
          \textbf{a} Energy and \textbf{b} force predictions compared with DFT reference values. 
          The energy and force RMSEs are $4.35~\mathrm{meV/atom}$ and $57.46~\mathrm{meV/\text{\AA}}$, respectively. 
           }
  \label{fig:nep_acc}
\end{figure}

The trained model reaches root-mean-square errors of $4.35~\mathrm{meV/atom}$ for energies and $57.46~\mathrm{meV/\text{\AA}}$ for forces relative to DFT (Fig.~\ref{fig:nep_acc}a,b).
It also reproduces key benchmark properties of $\alpha$-Fe and Fe-H systems, including elastic constants, surface energies, unstable stacking-fault energies, hydrogen diffusion, hydrogen adsorption, vacancy-H interactions, and dislocation-related energetics (see Table~\ref{tab:nep_benchmark} and Supplementary Information Section~S1).
For the present fracture problem, two additional validation points are central.
First, traction-separation tests in the Supplementary Fig.~S16-S18 capture hydrogen-induced weakening of separation resistance for the relevant low-index surfaces.
Second, the GPUMD implementation provides the efficiency required for three-dimensional crack models: on the benchmark shown in Supplementary Fig.~S19, the present NEP is approximately $600\times$ faster than the n2p2 implementation in LAMMPS on 72 CPU cores and approximately $100\times$ faster than the DeepMD implementation in LAMMPS on one NVIDIA H100 GPU.

\begin{table*}[htbp]
\centering
\caption{Benchmark properties for the $\alpha$-Fe/H NEP. The table compares the present NEP with DeepMD, n2p2, DFT, and available experimental data.}
\resizebox{\textwidth}{!}{%
\begin{tabular}{lllllll}
\toprule
Properties & Type & NEP & deepMD \cite{zhang2024highly} & n2p2 \cite{meng2021general} & DFT & Expt. \\
\midrule
Lattice constant ($\text{\AA}$) & $a_0$ & 2.835 & 2.834 & 2.83 & 2.83 & 2.855 \cite{basinski1955lattice} \\
Elastic constants (GPa) & $C_{11}, C_{12}, C_{44}$ & 250.7, 123.8, 98.7 & 280, 128, 104 & 296, 147, 96 & 297, 151, 105 & 239.5, 135.7, 120.7 \cite{adams2006elastic} \\
\multirow{4}{*}{Surface energy (J/m$^2$)} & (100) & 2.502 & 2.501 & 2.479 & 2.488 & \\
& (110) & 2.356 & 2.461 & 2.436 & 2.449 & \\
& (111) & 2.707 & 2.686 & 2.695 & 2.691 & \\
& (112) & 2.599 & 2.601 & 2.586 & 2.575 & \\
Unstable SFE (J/m$^2$) & $(112)\langle 11\bar{1}\rangle$ & 1.11 & 1.02 & 1.17 & & \\
Bonding energy of H$_2$ molecule (eV) & & 4.50 & 4.50 & 4.54 & & \\
Diffusion barrier of H atom (eV) & T-site to T-site & 0.091 & 0.099 & 0.108 & 0.092 & \\
\multirow{2}{*}{Trapping energies of H at vacancy (eV)} & 1st H & 0.640 & 0.713 & 0.592 & 0.616 \cite{hayward2013interplay} & \\
& 6th H & 0.078 & 0.035 & 0.046 & & \\
Adsorption energy of 1H atom on surface & (100) & 0.568 & 0.490 & 0.452 & 0.46 \cite{wang2014hydrogen} & \\
Vacancy formation energy (eV) & monovacancy & 2.293 & 2.327 & 2.203 & 2.223 & 2.0 \cite{de1983positron} \\
\multirow{3}{*}{Energy of dislocation relative to easy core (meV/$b$)} & Hard core & 28.9 & 51.5 & 47.4 & 39.3 \cite{itakura2012first}, 33.2 \cite{dezerald2014ab}, 57.9 \cite{wakeda2017chemical} & \\
& Split core & 108.35 & 69.9 & 82.3 & 108 \cite{itakura2012first}, 87.9 \cite{dezerald2014ab}, 110.3 \cite{wakeda2017chemical} & \\
& Saddle point & 29 & 34.8 & 38.2 & 37.9 \cite{itakura2012first}, 34.9 \cite{dezerald2014ab}, 49.2 \cite{wakeda2017chemical} & \\
\multirow{2}{*}{Kink nucleation energy (eV)} & Pure Fe & 0.74 & 0.7 & 0.7 & & \\
& With H & 0.52 & 0.62 & 0.67 & & \\
\bottomrule
\end{tabular}%
}
\label{tab:nep_benchmark}
\end{table*}

\FloatBarrier
\section{Results and Discussion}
\label{sec:results_discussion}
The central question is whether hydrogen provides an atomistic basis for a transgranular crack-plane transition from $\{100\}$ cleavage in hydrogen-free Fe toward $\{110\}$ cleavage under hydrogen charging.
Such a transition requires two conditions to be met simultaneously.
First, the selected brittle path must belong to a cleavage-plane family with a sufficiently low cleavage threshold; second, at that crack tip, cleavage must compete successfully against dislocation emission.
We therefore present the direct MD evidence first and then interpret it using Griffith cleavage thresholds and Rice-type emission-resistance descriptors; the former ranks competing cleavage paths, whereas the latter is used only as an energetic indicator of how hydrogen shifts the cleavage-emission balance.

Throughout this section, $\KMD$ denotes the local mode-I stress intensity factor (SIF) fitted from the MD stress field immediately before the first irreversible crack-tip event (see Methods).  
This direct MD observation determines whether the event is cleavage, dislocation emission, or mixed. 
The quantities $\KIG$ and $\KIe$ are instead idealized sharp-crack energetic quantities defined from the Griffith cleavage criterion and the Rice-type dislocation-emission criterion, respectively (see Methods). 
In particular, $\KIe$ should not be interpreted as the exact mode-I SIF at which the present thermostatted, slightly blunted, three-dimensional MD crack emits a dislocation. 
We therefore use $\KIe$ and $D=\KIe/\KIG$ as comparative indicators of hydrogen-induced trends in the cleavage-emission balance, while the actual event selection is determined directly from MD.

\begin{table}[H]
\centering
\caption{Consolidated fracture-selection metrics from constant-chemical-potential atomistic simulations and energetic analyses. The bulk hydrogen concentration $\cH$ is the equilibrium concentration corresponding to the imposed chemical potential $\muH$ in a bulk Fe reference. $\KMD$ is the local SIF fitted from the MD stress field at the first irreversible crack-tip event. $\gamma_s$ and $\KIG$ quantify cleavage-plane-family selection, while the active slip system, $\gamma_{\mathrm{usf}}$, $\KIe$, and $D=\KIe/\KIG$ provide a Rice-type energetic descriptor of the cleavage-emission balance. The values of $\KIe$ and $D$ are used to compare trends.}
\scriptsize
{
\resizebox{\textwidth}{!}{%
\begin{tabular}{lcccccccccc}
\toprule
Crack system & $\muH$ & Bulk $\cH$ & $\KMD$ & Fracture & $\gamma_s$ & $\KIG$ & Active slip & $\gamma_{\mathrm{usf}}$ & $\KIe$ & $D$ \\
 & (eV) & (appm) & (MPa$\cdot\mathrm{m}^{1/2}$) & mode & (J/m$^2$) & (MPa$\cdot\mathrm{m}^{1/2}$) & system & (J/m$^2$) & (MPa$\cdot\mathrm{m}^{1/2}$) & \\
\midrule
\multirow{4}{*}{$(100)[010]$} & $-\infty$ & 0 & 0.97 & Cleavage & 2.39 & 0.98 & $\{110\}\langle111\rangle$ & 0.99 & 1.75 & 1.79 \\
 & $-2.40$ & 0.0316 & 0.94 & Cleavage & 1.87 & 0.86 & $\{110\}\langle111\rangle$ & 0.99 & 1.75 & 2.03 \\
 & $-2.35$ & 0.331 & 0.92 & Cleavage & 1.77 & 0.84 & $\{110\}\langle111\rangle$ & 0.98 & 1.74 & 2.08 \\
 & $-2.30$ & 3.49 & 0.88 & Cleavage & 1.67 & 0.82 & $\{110\}\langle111\rangle$ & 0.96 & 1.73 & 2.12 \\
\addlinespace
\multirow{4}{*}{$(100)[011]$} & $-\infty$ & 0 & 0.98 & Cleavage & 2.39 & 0.99 & $\{112\}\langle111\rangle$ & 1.12 & 2.00 & 2.02 \\
 & $-2.40$ & 0.0316 & 0.94 & Cleavage & 1.87 & 0.88 & $\{112\}\langle111\rangle$ & 1.12 & 2.00 & 2.28 \\
 & $-2.35$ & 0.331 & 0.91 & Cleavage & 1.77 & 0.86 & $\{112\}\langle111\rangle$ & 1.12 & 2.00 & 2.34 \\
 & $-2.30$ & 3.49 & 0.89 & Cleavage & 1.67 & 0.83 & $\{112\}\langle111\rangle$ & 1.10 & 1.99 & 2.39 \\
\addlinespace
\multirow{4}{*}{$(110)[001]$} & $-\infty$ & 0 & 1.02 & Cleavage & 2.26 & 0.95 & $\{110\}\langle111\rangle$ & 0.99 & 1.60 & 1.69 \\
 & $-2.40$ & 0.0316 & 0.97 & Cleavage & 1.43 & 0.76 & $\{110\}\langle111\rangle$ & 0.99 & 1.60 & 2.12 \\
 & $-2.35$ & 0.331 & 0.92 & Cleavage & 1.29 & 0.72 & $\{110\}\langle111\rangle$ & 0.98 & 1.60 & 2.23 \\
 & $-2.30$ & 3.49 & 0.89 & Cleavage & 1.15 & 0.68 & $\{110\}\langle111\rangle$ & 0.96 & 1.58 & 2.34 \\
\addlinespace
\multirow{4}{*}{$(110)[1\bar{1}0]$} & $-\infty$ & 0 & 1.02 & Emission & 2.26 & 1.00 & $\{112\}\langle111\rangle$ & 1.12 & 1.36 & 1.36 \\
 & $-2.40$ & 0.0316 & 0.97 & Emission & 1.43 & 0.80 & $\{112\}\langle111\rangle$ & 1.12 & 1.36 & 1.70 \\
 & $-2.35$ & 0.331 & 0.94 & Mixed & 1.29 & 0.76 & $\{112\}\langle111\rangle$ & 1.12 & 1.36 & 1.79 \\
 & $-2.30$ & 3.49 & 0.88 & Cleavage & 1.15 & 0.72 & $\{112\}\langle111\rangle$ & 1.10 & 1.35 & 1.88 \\
\bottomrule
\end{tabular}%
}
}
\label{tab:md}
\end{table}

\subsection{Direct atomistic simulations identify the controlling crack system}
\begin{figure}[htbp]
\centering
\includegraphics[width=0.80\textwidth,height=0.84\textheight,keepaspectratio]{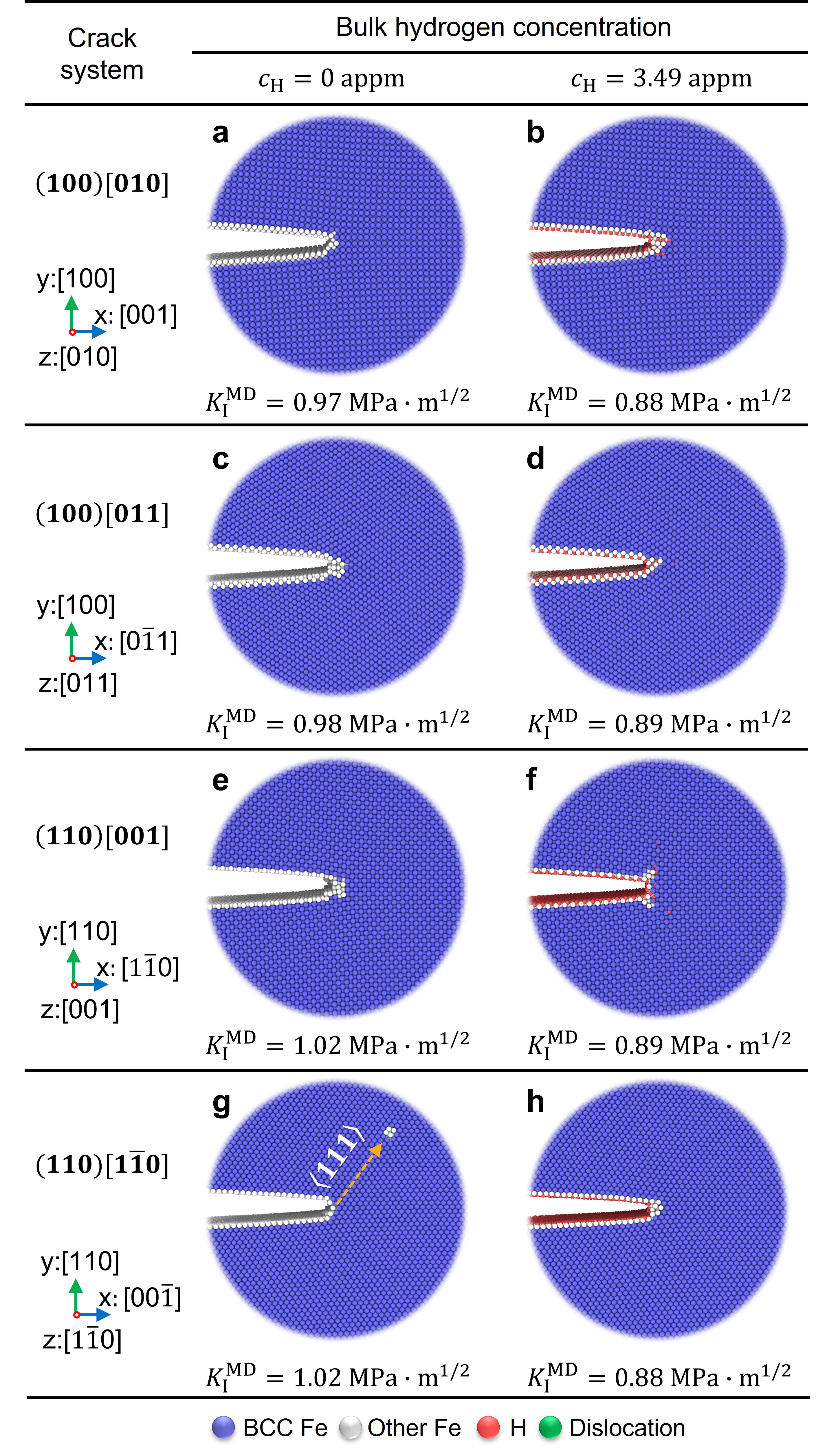}
\caption{Comparison of fracture responses for the four crack systems in pure Fe ($\cH=0$) and at the highest hydrogen condition ($\cH=3.49$~appm). 
         The $(100)[010]$, $(100)[011]$, and $(110)[001]$ systems remain cleavage-dominated, whereas the controlling $(110)[1\bar{1}0]$ system changes from dislocation emission in pure Fe to cleavage-dominated behavior under hydrogen charging. 
         The labels report $\KMD$, fitting from the MD stress field at the first irreversible event. 
         Full loading maps and trajectories are provided in Supplementary Fig.~S20 and Fig.~S21 and Supplementary Movies~1--4.
         }
\label{fig:4sys_map}
\end{figure}

Table~\ref{tab:md} summarizes the direct atomistic simulations for the four crack systems under four hydrogen conditions, while the same table also lists the energetic quantities used below to interpret cleavage selection and cleavage-emission competition.
Figure~\ref{fig:4sys_map} compares the direct MD observations for pure Fe and the highest hydrogen condition, $\cH=3.49$~appm. 
The central observation is that the $(100)[010]$, $(100)[011]$, and $(110)[001]$ cracks remain cleavage-dominated in both cases, whereas only $(110)[1\bar{1}0]$ changes its first irreversible event from dislocation emission in pure Fe to cleavage at high hydrogen content. 
The full 16-case maps of first irreversible event and dominant event are provided in Supplementary information section~S3, and the corresponding loading sequences are provided as Supplementary Movies~1--4.

Hydrogen lowers the MD-derived onset SIF in every crack system.
For $(100)[010]$, $(100)[011]$, and $(110)[001]$, this reduction mainly facilitates already selected cleavage paths and decorates the newly created crack surfaces with hydrogen. 
The $(110)[1\bar{1}0]$ crack behaves differently: in pure Fe, it is protected by dislocation emission, whereas under hydrogen charging it evolves through mixed cleavage-and-emission behavior and finally becomes cleavage-dominated.
Thus the central transition is not simply a monotonic reduction in fracture resistance; it is a change in the ordering of competing crack-tip pathways.

\begin{figure}[htbp]
\centering
\includegraphics[width=0.82\textwidth,height=0.88\textheight,keepaspectratio]{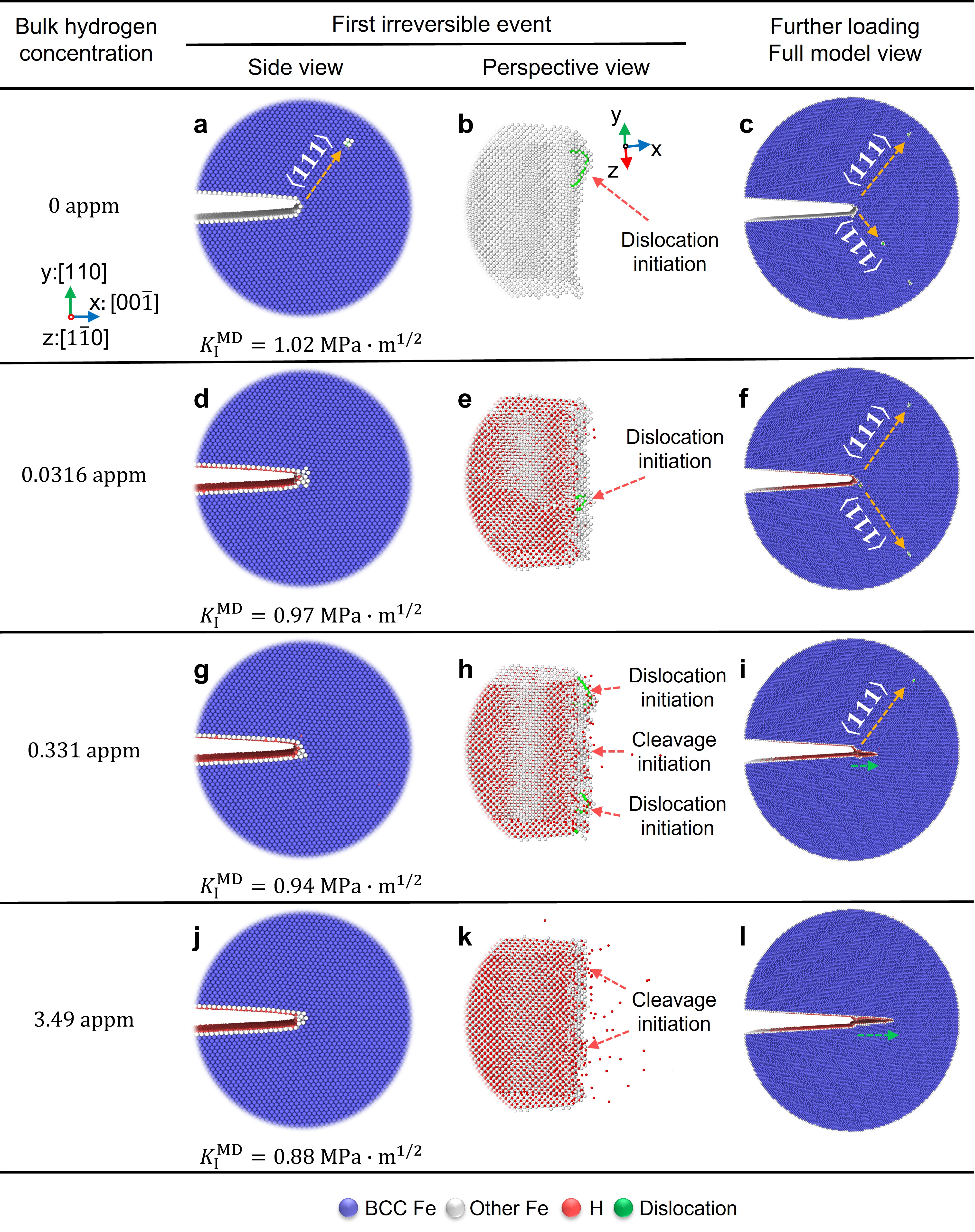}
\caption{Hydrogen-dependent evolution of the controlling $(110)[1\bar{1}0]$ crack. 
         Rows correspond to increasing bulk hydrogen concentration, and columns show the side view of the first irreversible event, the corresponding near-tip perspective view, and the dominant response after further loading to $\Kapp=1.80~\mathrm{MPa}\cdot\mathrm{m}^{1/2}$. 
         Pure Fe and the lowest hydrogen condition show dislocation emission, the intermediate condition shows mixed emission and cleavage, and the highest hydrogen condition shows cleavage-dominated fracture.
         }
\label{fig:transition_map}
\end{figure}

Figure~\ref{fig:transition_map} expands the controlling $(110)[1\bar{1}0]$ case identified in Fig.~\ref{fig:4sys_map}. 
The sequence shows that the change does not occur abruptly in a single step: the crack evolves from emission-dominated behavior in pure Fe and at low hydrogen content, through a mixed regime in which dislocation embryos and cleavage embryos coexist, to a cleavage-dominated response at the highest hydrogen content.
This direct atomistic evidence motivates the two energetic questions addressed below: why the $\{110\}$ cleavage-plane family becomes favorable for cleavage and why emission no longer shields the controlling $\{110\}$ crack effectively.

\FloatBarrier
\subsection{Hydrogen inverts the cleavage-plane-family ranking}
A crack system can become the preferred brittle path only if its cleavage threshold is competitive with those of other possible systems.
For the proposed crack-plane transition, the relevant comparison is not a single arbitrarily chosen crack direction but the relative behavior of the $\{100\}$ and $\{110\}$ crystallographic cleavage-plane families.
Using the thermodynamic integration and grand-canonical relations described in Methods, we evaluated the hydrogen-dependent surface free energies and the corresponding Griffith thresholds for all four crack systems.

\begin{figure}[htbp]
\centering
\includegraphics[width=0.95\textwidth]{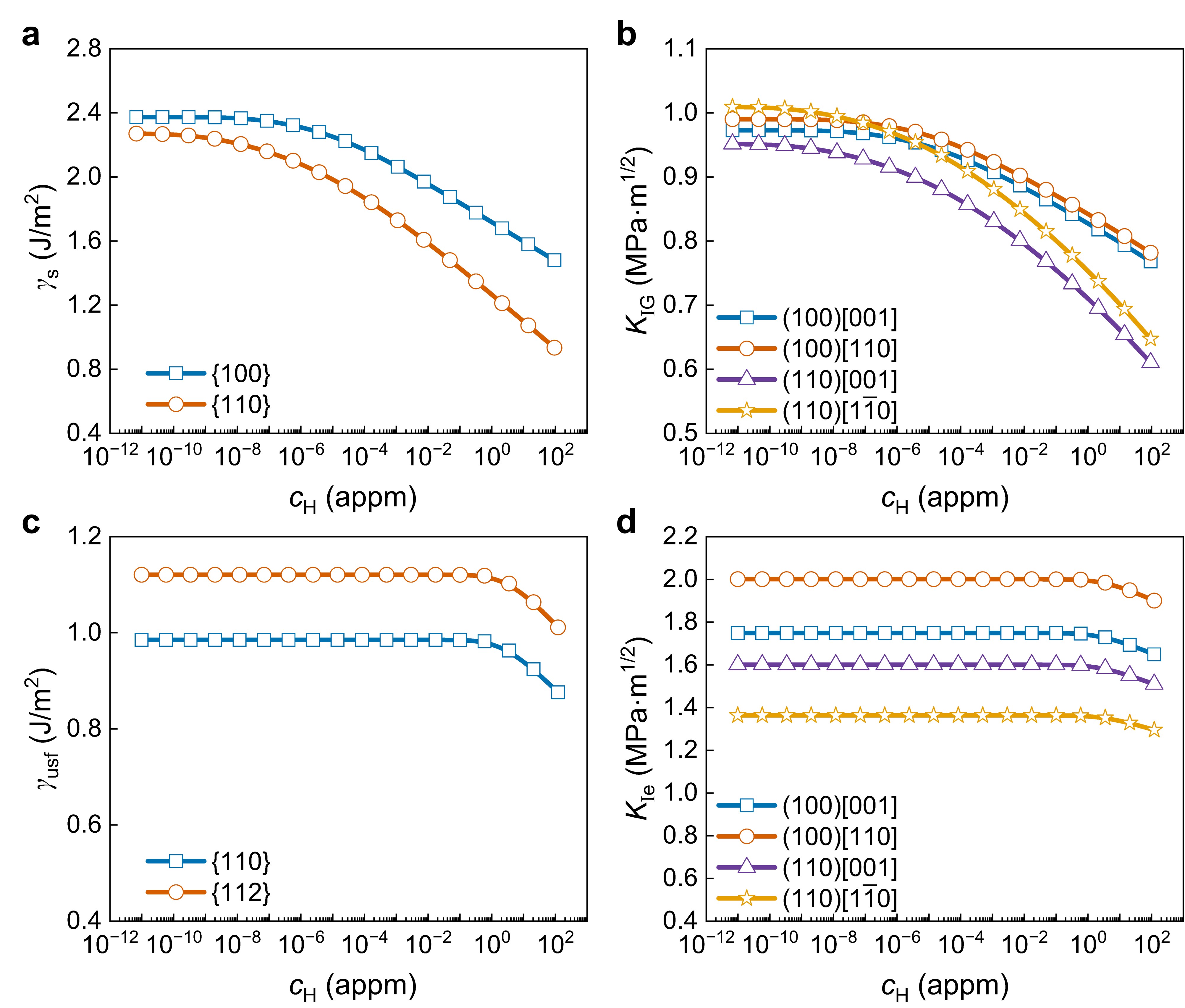}
\caption{Hydrogen dependence of the energetic quantities entering the Griffith/Rice-type comparison. 
        \textbf{a} Surface free energies of the $\{100\}$ and $\{110\}$ cleavage-plane families. 
        \textbf{b} Corresponding Griffith cleavage thresholds $\KIG$ for the four crack systems. 
        \textbf{c} Unstable stacking-fault energies of the slip families entering the Rice-type emission analysis. 
        \textbf{d} Corresponding Rice-type emission-resistance estimates $\KIe$. The bulk hydrogen concentration $\cH$ is the equilibrium concentration associated with the imposed chemical potential in a bulk Fe reference.
        }
\label{fig:thresholds}
\end{figure}

Hydrogen lowers the surface free energy of both the $\{100\}$ and $\{110\}$ cleavage-plane families, but the reduction is substantially stronger for $\{110\}$ (Fig.~\ref{fig:thresholds}a and Table~\ref{tab:md}).
Consequently, the Griffith threshold $\KIG$ decreases more rapidly for $\{110\}$ than for $\{100\}$ (Fig.~\ref{fig:thresholds}b).
The ratio plot $R$ in Fig.~\ref{fig:fig_mechanism}a makes the cleavage-plane-family crossover explicit: with increasing hydrogen content, the $\{110\}$ cleavage thresholds progressively undercut the corresponding $\{100\}$ thresholds.
This crossover supplies the first requirement for a crack-plane transition, namely that hydrogen makes $\{110\}$ cleavage competitive with, and eventually more favorable than, $\{100\}$ cleavage at the cleavage-plane-family level.

\FloatBarrier
\subsection{Hydrogen weakens the emission shield on the controlling $(110)[1\bar{1}0]$ crack}
\begin{figure}[htbp]
\centering
\includegraphics[width=0.95\textwidth]{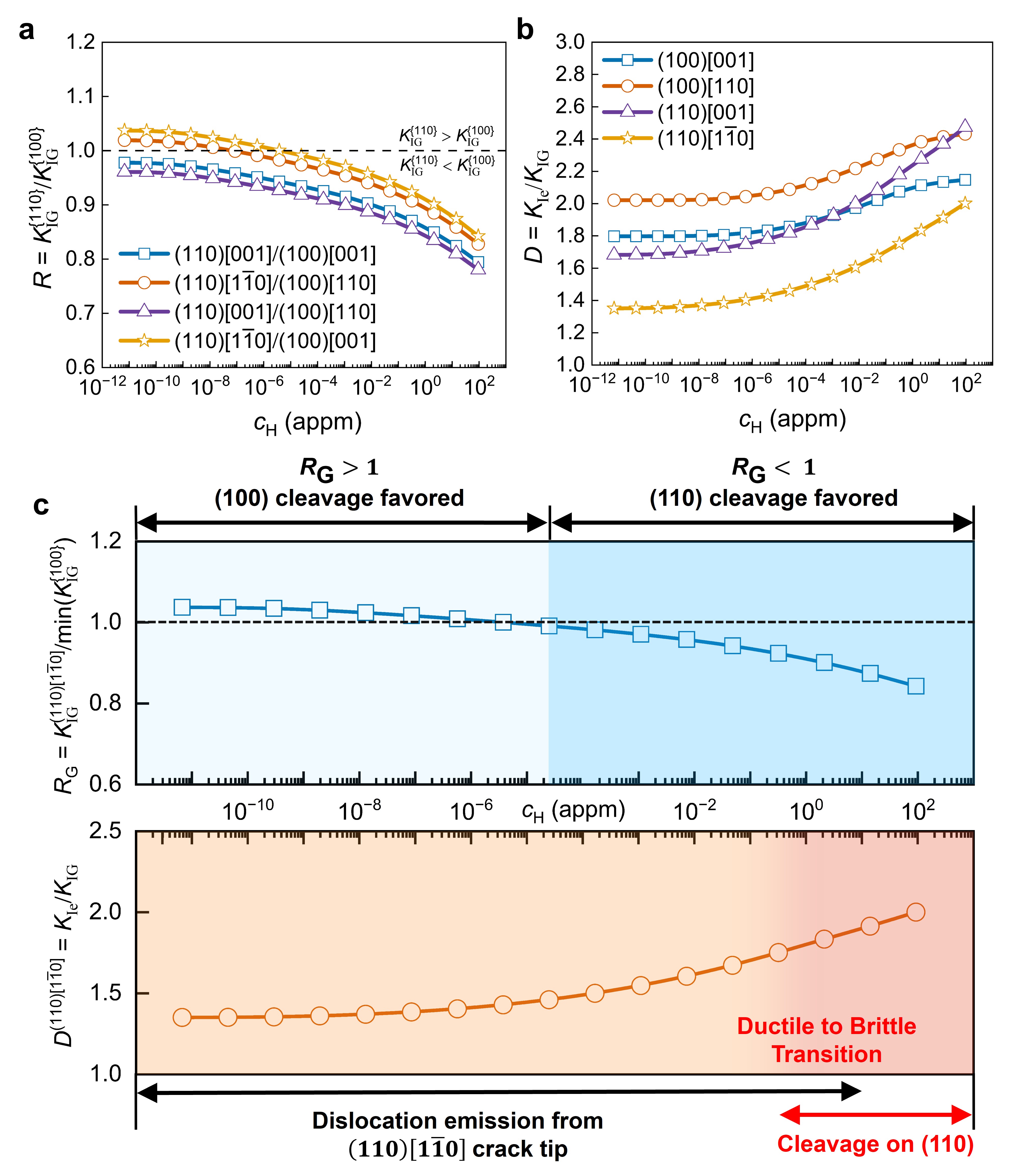}
\caption{Crossover criteria for the crack-plane transition. 
        \textbf{a} Ratios of $\{110\}$ to $\{100\}$ Griffith cleavage thresholds for different crack-front direction pairings; crossing below basis of cleavage-family selection indicates that the $\{110\}$ cleavage-plane family becomes favored relative to $\{100\}$. 
        \textbf{b} Rice-type index $D=\KIe/\KIG$ for the four crack systems; increasing $D$ indicates a trend toward greater cleavage competitiveness relative to dislocation emission, but $D$ is not used as an absolute MD event criterion. 
        \textbf{c} Two-step physical basis for the crack-plane transition. 
        Compact crossover map using the representation in \textbf{a} and \textbf{b}, shows the cleavage-selection ratio $R_{\rm G}=\KIG^{(110)[1\bar{1}0]}/\min(\KIG^{\{100\}})$ together with local mode transition $D$ for $(110)[1\bar{1}0]$ crack. 
        Crossover of $R_{\rm G}$ (horizontal grey dash line) indicates change in cleavage preference from $(110)$ cracks to $(110)[1\bar{1}0]$ crack.
        The rapid increase in $D$, together with the direct MD observations, local mode transition from dislocation emission to cleavage.
        Two criteria rationalize the associated conditions for cleavage plane transition toward the $\{110\}$ family.
        }
\label{fig:fig_mechanism}
\end{figure}

A low cleavage threshold alone does not guarantee brittle fracture, because dislocation emission can relax the crack-tip stress field and suppress cleavage.
The second requirement for the crack-plane transition is therefore local mode selection: on the controlling $\{110\}$ crack, cleavage must become more competitive relative to emission.
This competition is most important for $(110)[1\bar{1}0]$, the only system in which the direct atomistic response changes from dislocation emission to cleavage under hydrogen charging.
The Rice-type emission-resistance estimate $\KIe$ depends on the unstable stacking-fault energy of the active slip system and on anisotropic geometric factors (Methods). 
Hydrogen lowers the unstable stacking-fault energy, and therefore also lowers this Rice-type estimate (Fig.~\ref{fig:thresholds}c,d), but the decrease in the Griffith cleavage threshold is stronger for the controlling $(110)[1\bar{1}0]$ crack (Table~\ref{tab:md}). 
Although the raw unstable stacking-fault energy of the $\{112\}$ branch is higher than that of the $\{110\}$ branch, it is retained in the analysis because the active emission path is selected by the full Rice-type expression, which combines $\gamma_{\mathrm{usf}}$ with the anisotropic geometric factors for each crack orientation. 
The resulting $\KIe$ and $D=\KIe/\KIG$ are therefore not absolute brittle/ductile criteria for the present blunted, finite-temperature, three-dimensional crack geometry. 
This point is evident from the pure-Fe $(110)[1\bar{1}0]$ case, for which $D>1$ while MD directly shows dislocation emission. 
Their significance here is instead comparative: $D$ increases from $1.35$ in pure Fe to $1.88$ at $\cH=3.49$ appm (Fig.~\ref{fig:fig_mechanism}b), indicating that hydrogen shifts the energetic balance toward cleavage for the same crack system that changes from emission to mixed behavior and finally to cleavage in the direct MD simulations. 
A simple MD-normalized Rice-type estimate gives the same trend for the controlling crack (Supplementary Information Section~S4), but this normalization is used only as a robustness check and not as a definition of a true emission toughness. 
Thus, within the Griffith/Rice-type comparison, hydrogen does not merely soften the crack tip; it lowers the cleavage resistance more rapidly than the Rice-type emission-resistance estimate, consistent with a weakened dislocation-emission shield.
The compact map in Fig.~\ref{fig:fig_mechanism}c summarizes the two criteria together using the earlier crossover representation, linking the decreasing cleavage-selection ratio to the increasing emission-cleavage index.

\FloatBarrier
\subsection{A two-step physical basis for the crack-plane transition}
The simulations and energetic analyses together show that the hydrogen-induced crack-plane transition is controlled by a coupled reordering of the fracture landscape rather than by a single threshold.
In hydrogen-free Fe, $\{100\}$ remains the favorable transgranular cleavage-plane family, while the controlling $(110)[1\bar{1}0]$ crack is protected from cleavage by dislocation emission.
Under hydrogen charging, the $\{110\}$ cleavage resistance decreases more rapidly than the $\{100\}$ resistance, and the Rice-type descriptor for $(110)[1\bar{1}0]$ shifts toward cleavage because the surface-energy-controlled cleavage resistance decreases faster than the unstable-stacking-fault-controlled emission-resistance estimate.
When these two changes occur together, the preferred transgranular brittle fracture plane changes from $\{100\}$ toward $\{110\}$ as summarized in Fig.~\ref{fig:fig_mechanism}c.
The compact map in Fig.~\ref{fig:fig_mechanism}c shows that this energetic interpretation is consistent with the MD observations: the cleavage-plane-family crossover occurs as the $(110)[1\bar{1}0]$ response evolves from emission to mixed behavior and finally to cleavage.

\FloatBarrier
\section{Conclusion}
\label{sec:conclusion}
We have developed and validated an efficient NEP for the $\alpha$-Fe/H system and used it to perform chemically open, three-dimensional crack simulations at fixed hydrogen chemical potential.
The simulations provide atomistic evidence for a hydrogen-induced transgranular crack-plane transition that had been suggested by experiments but lacked a direct atomistic basis.
The direct comparison first isolates the controlling crack system: $(100)[010]$, $(100)[011]$, and $(110)[001]$ remain cleavage-dominated under hydrogen charging, whereas $(110)[1\bar{1}0]$ changes from dislocation emission in pure Fe to cleavage at elevated hydrogen chemical potential.
The detailed loading sequence for this controlling system further shows that the change proceeds through a mixed regime in which emission and cleavage coexist locally along the crack front.

The physical basis is twofold.
Hydrogen first changes the cleavage-plane-family ranking by lowering the $\{110\}$ Griffith threshold more strongly than the $\{100\}$ threshold.
It then shifts the Rice-type cleavage-emission descriptor for $(110)[1\bar{1}0]$ toward cleavage because the surface-free-energy reduction lowers the cleavage threshold faster than the unstable-stacking-fault-energy reduction lowers the emission-resistance estimate.
These coupled effects explain why hydrogen can favor transgranular cleavage on $\{110\}$ rather than simply reducing fracture resistance on the original $\{100\}$ path.
More broadly, the results show that environmentally assisted fracture should be treated as a chemically open atomistic problem in which the near-tip hydrogen state is solved under a specified chemical potential. This perspective is relevant for engineering assessments because it links hydrogen exposure conditions to the microscopic crack path and to the extent of dislocation shielding, rather than only to a scalar toughness reduction.

\FloatBarrier
\section{Methods}
\label{sec:methods}
\subsection{$\alpha$-Fe/H NEP framework and training}
The $\alpha$-Fe/H potential was developed within the NEP framework implemented in GPUMD \cite{fan2022gpumd,xu2025gpumd}.
In NEP, the total energy is written as a sum of site energies, and the site energy of atom $i$ is represented as a function of its descriptor vector, parameterized by a feedforward neural network with a single hidden layer containing $N_{\mathrm{neu}}$ neurons:

\begin{equation}
    U_i(\mathbf{q}^i) = U_i\left(\{q_\nu^i\}_{\nu=1}^{N_{\mathrm{des}}}\right),
\end{equation}

where $N_{\mathrm{des}}$ is the number of descriptor components.
The descriptor vector contains radial and angular components.
The radial component is

\begin{equation}
    q_n^i=\sum_{j\ne i} g_n(r_{ij}), \qquad 0\le n\le n_{\max}^{\mathrm R},
\end{equation}

where the summation is over neighbors within the radial cutoff and $g_n(r_{ij})$ is the radial basis function of the pair distance $r_{ij}$.
The three-body and four-body angular components are

\begin{equation}
    q_{nl}^{i}=\sum_{m=-l}^{l}(-1)^m A_{nlm}^{i} A_{nl(-m)}^{i},
\end{equation}

\begin{equation}
    q_{nl_1l_2l_3}^{i}=\sum_{m_1=-l_1}^{l_1}\sum_{m_2=-l_2}^{l_2}\sum_{m_3=-l_3}^{l_3}
    \begin{pmatrix}l_1&l_2&l_3\\m_1&m_2&m_3\end{pmatrix}
    A_{nl_1m_1}^{i}A_{nl_2m_2}^{i}A_{nl_3m_3}^{i},
\end{equation}

with

\begin{equation}
    A_{nlm}^{i}=\sum_{j\ne i}g_n(r_{ij})Y_{lm}(\theta_{ij},\phi_{ij}).
\end{equation}

Here $Y_{lm}$ are spherical harmonics, and $\theta_{ij}$ and $\phi_{ij}$ are the polar and azimuthal angles of the vector from atom $i$ to atom $j$.
The training hyperparameters are provided in Supplementary Information Section~S1.

\subsection{Anisotropic linear elastic fracture mechanics}
Because $\alpha$-Fe is elastically anisotropic, we used anisotropic linear elastic fracture mechanics (ALEFM) to construct the mode-I boundary displacement field and to evaluate Griffith and Rice-type thresholds.
Under plane strain and infinitesimal deformation, the Lekhnitskii formalism gives the governing equation for the Airy stress function $U$ as

\begin{equation}
    b_{11}\frac{\partial^4 U}{\partial y^4}+b_{22}\frac{\partial^4 U}{\partial x^4}+(2b_{12}+b_{66})\frac{\partial^4 U}{\partial x^2\partial y^2}
    -2b_{16}\frac{\partial^4 U}{\partial x\partial y^3}-2b_{26}\frac{\partial^4 U}{\partial x^3\partial y}=0,
\label{eq:governing}
\end{equation}

where $b_{ij}$ are the reduced elastic coefficients in the rotated crack coordinate system.
The associated characteristic equation is

\begin{equation}
    b_{11}\mu^4-2b_{16}\mu^3+(2b_{12}+b_{66})\mu^2-2b_{26}\mu+b_{22}=0.
\label{eq:characteristic}
\end{equation}

The two roots with positive imaginary parts are denoted by $s_1$ and $s_2$.
For mode-I loading, the near-tip stress field is

\begin{equation}
\sigma_{xx}=\frac{K_{\mathrm I}}{\sqrt{2\pi r}}\mathrm{Re}\left[\frac{s_1s_2}{s_1-s_2}\left(\frac{s_2}{\sqrt{\cos\theta+s_2\sin\theta}}-\frac{s_1}{\sqrt{\cos\theta+s_1\sin\theta}}\right)\right],
\end{equation}
\begin{equation}
\sigma_{yy}=\frac{K_{\mathrm I}}{\sqrt{2\pi r}}\mathrm{Re}\left[\frac{1}{s_1-s_2}\left(\frac{s_1}{\sqrt{\cos\theta+s_2\sin\theta}}-\frac{s_2}{\sqrt{\cos\theta+s_1\sin\theta}}\right)\right],
\label{eq:sigma_yy}
\end{equation}
\begin{equation}
\sigma_{xy}=\frac{K_{\mathrm I}}{\sqrt{2\pi r}}\mathrm{Re}\left[\frac{s_1s_2}{s_1-s_2}\left(\frac{1}{\sqrt{\cos\theta+s_1\sin\theta}}-\frac{1}{\sqrt{\cos\theta+s_2\sin\theta}}\right)\right],
\end{equation}

where $r$ and $\theta$ are polar coordinates with respect to the crack tip and $K_{\mathrm I}$ is the mode-I SIF.
Detailed derivations are given in Refs.~\cite{stroh1958dislocations,sun2011fracture} and Supplementary Information Section~S5.

The energy release rate is written as

\begin{equation}
    G_{\mathrm I}=B K_{\mathrm I}^2,
\end{equation}

where the anisotropic elastic coefficient $B$ is

\begin{equation}
    B=\sqrt{\frac{b_{11}b_{22}}{2}\left(\frac{2b_{12}+b_{66}}{2b_{11}}+\sqrt{\frac{b_{22}}{b_{11}}}\right)}.
\label{eq:B}
\end{equation}

For cleavage, Griffith theory gives $G_{\mathrm I}=2\gamma_s$, leading to

\begin{equation}
    \KIG=\sqrt{\frac{2\gamma_s}{B}},
\label{eq:KIG}
\end{equation}

where $\gamma_s$ is the surface free energy of the cleavage plane.
For dislocation emission, the Rice crack-tip nucleation criterion and its anisotropic extension give a Rice-type emission-resistance estimate \cite{rice1992dislocation,sun1994dislocation}

\begin{equation}
    \KIe=\frac{\sqrt{\gamma_{\mathrm{usf}}\,o(\Phi,\theta)}}{F_{12}(\theta)},
\label{eq:KIe}
\end{equation}

where $\gamma_{\mathrm{usf}}$ is the unstable stacking-fault energy of the active slip plane, $o(\Phi,\theta)$ is an anisotropic elastic factor, and $F_{12}(\theta)$ is a geometric factor.
The angle $\theta$ is between the slip plane and crack plane, and $\Phi$ is between the slip direction and the vector perpendicular to the crack line on the slip plane (Fig.~\ref{fig:model}a).
Explicit expressions for $o(\Phi,\theta)$ and $F_{12}(\theta)$ are given in Supplementary Information Section~S5.
Equation~\eqref{eq:KIe} uses $\gamma_{\mathrm{usf}}$ from a generalized stacking-fault calculation together with anisotropic geometric factors. 
It does not contain the full three-dimensional dislocation-embryo shape, thermal activation, crack-tip core structure, or the local hydrogen configuration at the instant of emission. 
We therefore treat $\KIe$ as a Rice-type energetic descriptor, rather than as the exact MD emission SIF; the actual first event is determined directly from the atomistic simulations.

\subsection{Crack model and cracking simulations}
\begin{figure}[hbp]
  \centering
  \includegraphics[width=1.0\textwidth]{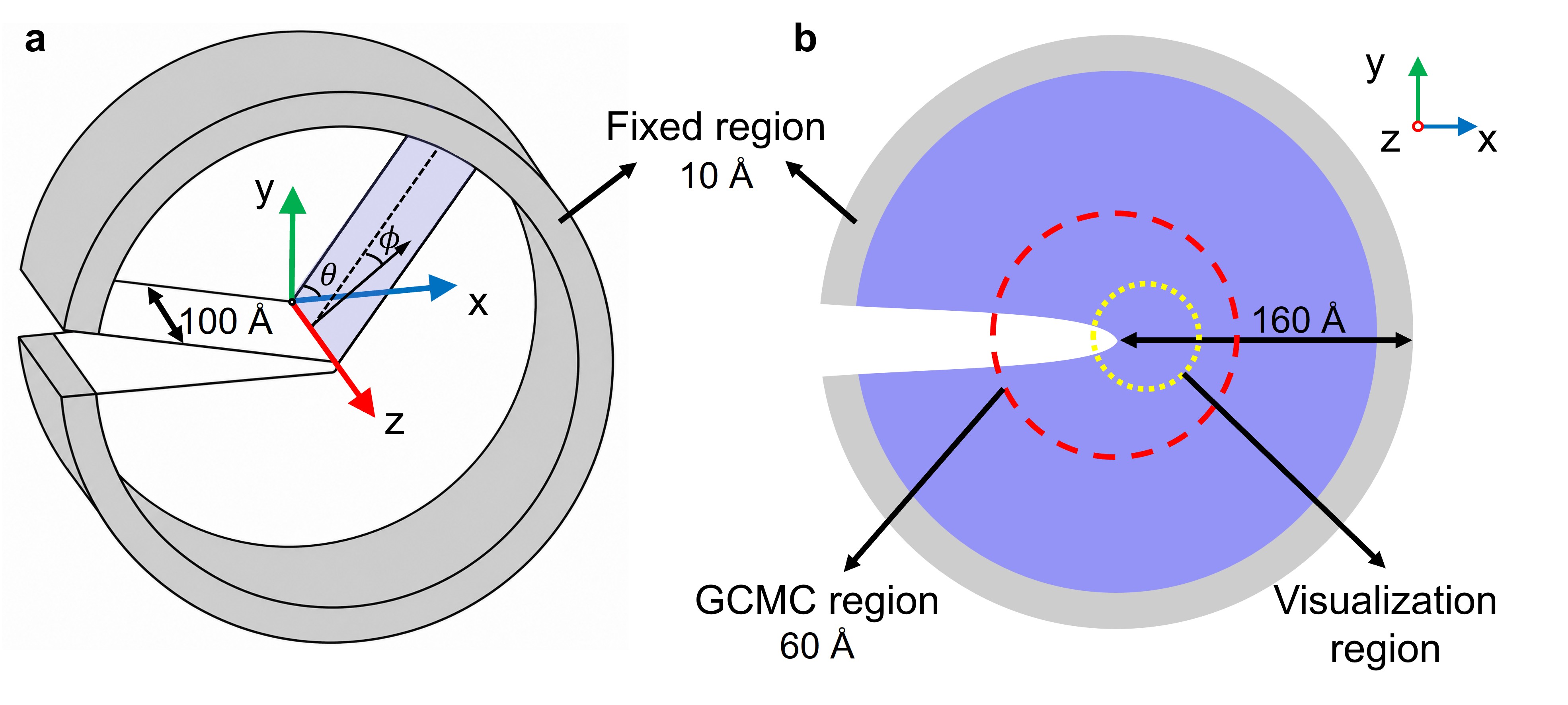}
  \caption{Crack geometry and constant-chemical-potential simulation setup. 
          \textbf{a} Cylindrical crack model used in the study. 
          A crack of length $L\simeq100~\text{\AA}$ is created along the $z$ axis. 
          The definitions of the slip plane and the angles $\theta$ and $\Phi$ used in the Rice type criterion are shown.
          \textbf{b} Loading and constant-chemical-potential simulation setup 
          Boundary atoms in the outer $10~\text{\AA}$ layer(grey) are displaced according to the ALEFM solution, while mobile atoms (purple) are equilibrated at 300 K after each loading increment. GCMC trial moves are performed within $r=60~\text{\AA}$ of the crack tip (red). Yellow region is used for virtualization of crack tip configuration. 
          }
  \label{fig:model}
\end{figure}

A cylindrical simulation cell containing a semi-infinite crack is constructed for each crack system (Fig.~\ref{fig:model}b).
The initial slit is introduced by removing two half atomic layers, creating a blunted-crack geometry frequently used in atomistic fracture simulations to avoid artificial bonding across crack faces \cite{andric2019atomistic,hiremath2024phosphorus}.
We therefore describe this geometry as a blunted crack rather than as an atomically sharp Griffith crack.
A size-convergence test (Supplementary Information Section~S6) is used to choose a cylinder radius of $R\simeq160~\text{\AA}$, and the crack length was $L\simeq100~\text{\AA}$.
Periodic boundary conditions are applied along the crack-front direction $z$, while the $x$ and $y$ directions have free surfaces; $x$ is the crack-propagation direction and $y$ is normal to the crack plane.

For mode-I loading, the displacement field derived from ALEFM is applied to atoms in an outer fixed layer of thickness $10~\text{\AA}$:

\begin{equation}
 u_x=K_{\mathrm I}\sqrt{\frac{2r}{\pi}}\mathrm{Re}\left[\frac{s_1p_2\sqrt{\cos\theta+s_2\sin\theta}-s_2p_1\sqrt{\cos\theta+s_1\sin\theta}}{s_1-s_2}\right],
\label{eq:ux}
\end{equation}
\begin{equation}
 u_y=K_{\mathrm I}\sqrt{\frac{2r}{\pi}}\mathrm{Re}\left[\frac{s_1q_2\sqrt{\cos\theta+s_2\sin\theta}-s_2q_1\sqrt{\cos\theta+s_1\sin\theta}}{s_1-s_2}\right],
\label{eq:uy}
\end{equation}

where $p_1$, $p_2$, $q_1$, and $q_2$ are elastic coefficients defined by the Stroh/Lekhnitskii solution and listed in Supplementary Information Section~S5.
Loading is increased in increments of $\Delta K_{\mathrm I}=0.01~\mathrm{MPa}\cdot\mathrm{m}^{1/2}$.
After each increment, mobile atoms are equilibrated in the $\mathrm{NVT}$ ensemble at $T=300~\mathrm K$ for 25 ps using a time step of 0.5 fs.

Because the initial crack is slightly blunted, $\Kapp$ should be interpreted as the externally imposed elastic boundary parameter, not as the local crack-tip driving force for an atomically sharp crack.
To compare crack-tip events across systems, we therefore define $\KMD$ as the local SIF fitted from the MD stress field immediately before the first irreversible crack-tip event.
Along the crack-propagation direction ahead of the crack tip, the mode-I opening stress reduces to

\begin{equation}
    \sigma_{yy}=\frac{K_{\mathrm I}}{\sqrt{2\pi d}},
\label{eq:sigma_fit}
\end{equation}

where $d$ is the distance ahead of the crack tip.
The atomic virial stress is averaged along the crack-front direction and fitted to Eq.~\ref{eq:sigma_fit} outside the immediate atomic core region, as detailed in Supplementary Information Section~S7.
The fitted value at the first irreversible event is reported as $\KMD$.

\subsection{Hybrid GCMC/MD simulations}
Hydrogen is introduced by a hybrid grand-canonical Monte Carlo/molecular dynamics (GCMC/MD) protocol.
The simulations are performed in the $\mu\mathrm{VT}$ ensemble at $T=300~\mathrm K$.
GCMC trial insertion and deletion moves are attempted every 10 MD steps, with 100 trial moves per GCMC step, using an in-house GCMC implementation for GPUMD.
The implementation is validated by comparing equilibrium bulk hydrogen concentrations with standard GCMC calculations in LAMMPS (Supplementary Information Section~S8).

For crack simulations, GCMC moves are restricted to a region within $r=60~\text{\AA}$ of the crack tip.
This accelerates equilibration of the near-tip hydrogen population and avoids artificial exchange with distant free surfaces.
For each chemical potential, hydrogen is first equilibrated in the initial crack model at $\Kapp=0.91~\mathrm{MPa}\cdot\mathrm{m}^{1/2}$ for 350 ps.
Subsequent loading is performed using the same hybrid GCMC/MD protocol so that the near-tip crack-surface region and crack tip inside the GCMC domain remains coupled to the hydrogen reservoir throughout the loading sequence.
The protocol determines the thermodynamic hydrogen occupancy associated with each imposed chemical potential and loading state; it is not intended to model the rate at which hydrogen is delivered to a crack tip from a remote source.

\FloatBarrier
\subsection{Hydrogen-dependent surface and stacking-fault energetics}
Finite-temperature surface free energies are calculated by nonequilibrium thermodynamic integration from the solid to an Einstein crystal reference \cite{freitas2016nonequilibrium}.
At fixed hydrogen chemical potential, the surface free energy can be written in terms of the surface excess hydrogen population as
\begin{equation}
    \gamma_s(\muH)=\gamma_s^0-\frac{1}{2A}\int_{-\infty}^{\muH}\langle N_{\mathrm H}^{\mathrm{surf}}\rangle_{\mu'}\,d\mu',
\label{eq:gamma_s_mu}
\end{equation}
where $\gamma_s^0$ is the hydrogen-free surface free energy, $A$ is the surface area of one face, the factor of 2 accounts for the two surfaces in the slab, and $\langle N_{\mathrm H}^{\mathrm{surf}}\rangle_{\mu'}$ is the ensemble-averaged excess number of H atoms associated with the two surfaces at chemical potential $\mu'$.
An analogous expression is used for the unstable stacking-fault energy,
\begin{equation}
    \gamma_{\mathrm{usf}}(\muH)=\gamma_{\mathrm{usf}}^0-\frac{1}{A}\int_{-\infty}^{\muH}\langle N_{\mathrm H}^{\mathrm{usf}}\rangle_{\mu'}\,d\mu',
\label{eq:gamma_usf_mu}
\end{equation}
where $\langle N_{\mathrm H}^{\mathrm{usf}}\rangle_{\mu'}$ is the excess H population associated with the unstable stacking-fault configuration.
These thermodynamic relations ensure that $\gamma_s$ and $\gamma_{\mathrm{usf}}$ are evaluated under the same hydrogen chemical potential used in the crack simulations.

\section*{Author contributions}
All authors designed the research together. 
J.X., Z.Z., F.M. and S.Z. developed and validated the potential. 
J.X., S.S. and S.O. designed the simulations. 
J.X. conducted the simulations and theoretical calculations. 
J.X, Z.Z. and S.O. wrote the manuscript.  
All the authors commented on the manuscript.
\section*{Data availability}
The training command line is shared online: https://github.com/Ricky-Zhao/Neural-network-potential-for-the-binary-alpha-Fe-H-system. 
Other data used in this work are available from the authors upon request.

\section*{Code availability}
The constructed neural-network interatomic potential in this study is shared online: https://github.com/Ricky-Zhao/Neural-network-potential-for-the-binary-alpha-Fe-H-system

\section*{Competing interests}
The authors declare no competing interests.
\section*{Acknowledgements}
The authors gratefully acknowledge support from Nippon Steel Corporation. 
Z.Z. acknowledges support from Japan Society for the Promotion of Science (JSPS) Postdoctoral Fellowships for Research in Japan(Standard) (Fellowship ID: P25373). 
S.S. acknowledges support from JSPS KAKENHI Grant No. JP24K17170. 
S.Z. acknowledges support from JSPS KAKENHI Grant No. JP25K17509. 
S.O. acknowledges support from JSPS KAKENHI Grant Numbers JP23H00161, JP21K18675, and JP23K20037 and Ministry of Education, Culture, Sport, Science and Technology of Japan (MEXT) Programs, Grant Numbers JPMXP1122684766, JPMXP1020230325, and JPMXP1020230327. 
Part of this work was carried out using computational resources provided by large-scale computer systems at D3 Center, The University of Osaka, the Large-scale Parallel Computing Server at the Center for Computational Materials Science, Institute for Materials Research, Tohoku University, and the supercomputer system of The University of Tokyo through the HPCI System Research Project (Project ID: hp260145).
\printbibliography

\end{document}